\documentstyle[aps]{revtex}
  \advance\oddsidemargin  by -1in
  \advance\evensidemargin by -1in
\def\lsim{\mathrel{\rlap{\lower4pt\hbox{\hskip1pt$\sim$}}
    \raise1pt\hbox{$<$}}}         
\def\gsim{\mathrel{\rlap{\lower4pt\hbox{\hskip1pt$\sim$}}
    \raise1pt\hbox{$>$}}}         

\def\be{\begin{equation}}
\def\ee{\end{equation}}
\def\bq{\begin{eqnarray}}
\def\eq{\end{eqnarray}}

\newcommand{\doublespace}{                                                     
    \renewcommand{\baselinestretch}{1.5}\large\normalsize}
\catcode`\@=11
\def\eqalign#1{\null\,\vcenter{\openup\jot\m@th
\ialign{\strut\hfil$\displaystyle{##}$&$\displaystyle{{}##}$\hfil
     \crcr#1\crcr}}\,}
\catcode`\@=12
 
\begin{document}
\doublespace
\pagestyle{empty}
\hfill{

\hfill 

{\hfill ISN 96.93}

{\hfill August, 1996}
 
\bigskip
\vskip 12pt 
\centerline{\bf A UNIFIED PICTURE OF GLUEBALL CANDIDATES $f_0(1500)$ AND
$f_0(1700)$}
 
\vskip 36pt
\centerline{Marco Genovese \footnote{Supported by EU Contract ERBFMBICT
950427}}
\centerline{\small Institut des Sciences Nucl\'eaires}
\centerline{\small Universit\'e Joseph Fourier--IN2P3-CNRS}
\centerline{\small 53, avenue des Martyrs, F-38026 Grenoble Cedex,  
France}

\bigskip 
\centerline{\bf ABSTRACT}
\medskip 
A simple mixing scheme describing the $f_0(1500)$ and the $f_0(1700)$ as mixed
states of a $\bar s s$ meson and a digluonium is reconsidered at the light of
new experimental data.

\vspace{2.0cm}
 
In QCD, because of self interaction of gluons, one expects the existence
of particles composed of gluons, which are called glueballs.

Many years have passed since the original prediction of the existence of these
states \cite{gg}, but, albeit the existence of many different candidates,
up to now none of the many resonances has been assigned to a glueball.

Recently, two identifications of the lowest--lying scalar glueball have been
proposed: one claim \cite{Close} is that it corresponds to a resonance observed
at $1500$ MeV (which can be identified with the former $f_0(1590)$, see
\cite{PDB}) that does not fit the usual meson nonet; the second
\cite{Wein} identifies this state with the resonance observed at $1700$ MeV,
which was known as $\Theta(1700)$.  

Both the resonances have unusual decay properties for an ordinary $\bar q q$
meson, appear in gluon--rich channels 
and are in the mass region where lattice QCD predicts the existence of a
scalar glueball \cite{Wein,UKQCD}. Furthermore, for the $f_0(1700)$, Ref.
\cite{Wein} claims that the branching ratios are in agreement with a lattice
QCD calculation and that the mass practically coincides with the predicted one. 

However, it must be emphasized that all available lattice
calculations of the digluonium mass are made in the so--called quenched
approximation, namely neglecting the creation of $\bar q q$ pairs: this casts
some doubts about the reliability of these predictions, which can be considered
only as indicative. Also, the prediction of
branching ratios in lattice QCD has been questioned \cite{Martinelli}. 

Anyway, it is quite puzzling to observe in the region where the scalar digluonium
state is predicted two resonances not well fitting the usual meson nonets
and with unusual decay properties.

In our opinion a common origin should be
searched for both these states. Some years ago,
a phenomenological scheme was proposed, consisting of a mixing
between a glueball and a $\bar s s$ state \cite{io}. 
The improved
experimental data on these two resonances allow now quite a deeper
investigation. 
In this paper this mixing scheme is reconsidered
at the light of the new experimental inputs.

The mixing scheme is
\bq
  \mid f_{0}(1500) \rangle &= \cos\alpha \mid gg \rangle - 
\sin\alpha \mid \bar{s}s \rangle \nonumber \\
 \mid f_{0}(1700) \rangle &= \sin\alpha \mid gg \rangle 
+ \cos\alpha \mid \bar{s}s 
\rangle
\label{1}
\eq

Given that the main $f_{0}(1700)$ decay is in the $\bar{K}K$ channel, it is
assumed that the $\mid\bar{q}q \rangle$ 
($\bar{u}u$ and $\bar{d}d$) component is negligible \cite{io}. 

Of course, generally speaking, one could choose any arbitrary mixing between
the scalar digluonium, the light quark $\mid \bar  q q \rangle$ and the 
$\mid \bar s s \rangle$ states.
At the moment the experimental data do not yet allow to exclude other
choices, also because of the poor knowledge of the scalar nonet 
\cite{Penn,io2} compared to the other meson multiplets (it is worth to
remember that strong instantonic effects are expected in the scalar sector
\cite{Inst}).

For example, quite a different perspective
is adopted in \cite{Close}, based on
a mixing of the $0^{++}$ glueball 
with a light--quark system giving two physical states corresponding to the
$f_0(1500)$ and $f_0(1400)$. In \cite{Klempt} the $f_0(1500)$ derives
from a mixing of the $\mid \bar q  q \rangle$ and the $\mid \bar s s \rangle$
states. Finally, in \cite{Anis} an identification 
of $f_0(1500)$ with an unmixed glueball is
suggested.
               
In the following, besides the mixing hypothesis (\ref{1}), 
it is also assumed that the 
$f_{0}(1500)$ decouples from $\bar{K}K$, since no experimental 
evidence of this decay has been reported. This assumption permits to relate the 
decay amplitudes for the $gg$ and the $\bar s s$ components through
\bq  
&\langle K^{+} K^{-} \mid f_{0}(1500) \rangle = 0 =  \nonumber \\
&= \cos\alpha \langle K^{+} K^{-} \mid gg \rangle - \sin\alpha \langle K^{+} K^{-} 
\mid \bar{s}s \rangle
\label{2}
\eq

Of course this is quite an approximation, but it is the simplest one and
permits to obtain a large predictive power for the model. A small branching
ratio of $f_0(1500) \rightarrow K \bar K$ or some different phase between the
two amplitudes should not spoil the essence of the model. The agreement among
the results and the experimental data confirms that this ansatz is acceptable.

To obtain the mixing angle from  
\be
\langle K^{+} K^{-} \mid f_{0}(1700) \rangle = \csc\alpha \langle K^{+} K^{-} \mid gg 
\rangle \label{3}
\ee
one uses the isospin relations

\be
\Gamma(f \to \pi \pi ) = 3/2\times \Gamma(f \to \pi^{+} \pi^{-})
\ee
 
\be 
\Gamma( f \to KK ) = 2\times \Gamma(f \to K^{+} K^{-});
\ee
 
\noindent requests
$\langle \pi \pi \mid \bar{s}s \rangle = 0$ 
and assumes flavour independence
$$\langle K^{+} K^{-} \mid gg \rangle = \langle \pi^{+} \pi^{-} \mid gg \rangle .$$
Using the experimental values \cite{PDB} 
\bq
& B.R.[ f_{0}(1700) \to \pi \pi ] = 0.039 ^{+0.002}_{-0.024} \nonumber \\ 
& B.R.[ f_{0}(1700) \to \bar{K} K ] = 0.38 ^{+0.09}_{-0.19}  \label{4}
\eq   
from Eq. (\ref{3}), including the phase space factors $p$ (meson momentum), 
one gets
\bq
& \sin (\alpha) = 0.579 ^{+ 0.072} _{-0.095} & \nonumber \\
& \cos (\alpha) = 0.815 ^{+0.067} _{-0.051} &
\eq

From the mixing angle $\alpha$, one can immediately estimate the glueball and the 
scalar $\bar{s}s$ state masses \cite{io}, using the relations:
\bq
\pmatrix{\bar{s}s&\epsilon\cr
                 \epsilon&gg\cr} 
\pmatrix{-\sin\alpha\cr
         \cos\alpha} = 1.503 \pmatrix{-\sin\alpha\cr
                                     \cos\alpha} 
\eq
\bq
 \pmatrix{\bar{s}s&\epsilon\cr
                 \epsilon&gg\cr} 
\pmatrix{\cos\alpha\cr
         \sin\alpha} = 1.697 \pmatrix{\cos\alpha\cr
                                    \sin\alpha} 
\eq

\noindent where $gg$ denotes the digluonium mass, $\bar{s}s$ the quark state mass 
and $\epsilon$ the mixing parameter.   

This gives both  $\bar s s$ and $gg$ masses 
around $1.6$ GeV, the latter in agrement with lattice predictions 
\cite{UKQCD,Wein}.

The mixing angle gives also access to many branching
ratios. Using Ref. \cite{Close} and Eq. (\ref{3}) one gets
the following amplitudes relative to 
$\langle \pi \pi \mid gg \rangle $:
\be
\eqalign{
&\langle \pi \pi \mid  \bar s s \rangle = 0 \nonumber  \cr
&\langle K \bar K \mid \bar s s \rangle = R \cot (\alpha) \nonumber  \cr 
&\langle \eta \eta \mid \bar s s \rangle = 2 R \sin^2(\phi) \cot (\alpha) 
\nonumber  \cr 
&\langle \eta \eta' \mid  \bar s s \rangle = 
-2 R \cos(\phi) \sin(\phi) \cot (\alpha) \nonumber  \cr 
&\langle K \bar K \mid gg \rangle = R \nonumber  \cr
&\langle \eta \eta \mid gg \rangle = \cos^2(\phi)+ 
R^2 \sin^2(\phi) \nonumber  \cr 
&\langle \eta \eta' \mid gg \rangle = 
\cos(\phi) \sin(\phi) (1 - R^2) \nonumber \, , \cr
\label{5}}
\ee 
where $R = \langle  \bar s s \mid gg \rangle /
 \langle \bar q q \mid gg \rangle $ measurs the breaking of
$SU_f (3)$ in gluonium decays ($u$ and $d$ quarks are assumed to be equivalent). 
No flavour violation is considered for
the decay of quarkonium in pair of mesons (empirically this
violation is shown to be quite small for the well--established meson nonets
\cite{flavviol}).

\noindent In Eq. (\ref{5}), 
$\phi$ is the angle for the $\eta$ -- $\eta '$ mixing
\bq
&\eta = \cos (\phi)   | \bar q q  \rangle - \sin (\phi) |
\bar s s \rangle \nonumber  \\ 
&\eta ' = \sin (\phi)   | \bar q q \rangle + \cos (\phi) |
\bar s s \rangle
\label{6}
\eq
In the following, a value $\phi = 72 ^{o}$ \cite{cbarrel} is adopted; 
a possible
gluonic component of the $\eta$ and $\eta '$ (see for example \cite{GLP}
and references therein)
will not be considered here.  

Furthermore, considering that glueballs should, at least in a first 
approximation,  exhibit flavour democracy,  
flavour--independence for the $gg$ decays (i.e. $R=1$) as well is assumed.

Some predictions are listed here and compared with the data of Ref. \cite{PDB}.
The calculations include a sum over permutation and over the various charge
combinations with the appropriate weighting factors (4 for $ K \bar K$, 
3 for $\pi \pi$, 2 for $\eta \eta '$ and 1 for $\eta \eta$).
The theoretical errors only account for the uncertainty on the mixing angle.

\be
{\Gamma(f_0(1700) \rightarrow \eta \eta ) \over \Gamma(f_0(1700) 
\rightarrow \pi \pi)} = { p_{\eta} \over 3 p_{\pi}} \cdot \left[
1 + 2 \sin^2 (\phi) \cot ^2 (\alpha) \right ] ^2 =
5.43 ^{+3.0}_{-2.4}    \label{7}
\ee
in agreement with the experimental datum $4.6 ^{+2.9} _{-3.3}$;

\be
{\Gamma(f_0(1700) \rightarrow \eta \eta ) \over \Gamma(f_0(1700) 
\rightarrow K \bar K)} = { p_{\eta} \over 4 p_{K}} \cdot \left[
\sin^2 (\alpha) + 2 \sin^2 (\phi) \cos ^2 (\alpha) \right ] ^2 =
0.558 ^{+0.080}_{-0.061}    \label{7b}
\ee
in good agreement with the datum  $0.47 ^{+ 0.24} _{-0.35}$.

Similarly one finds:
\be
{\Gamma(f_0(1500) \rightarrow \pi^0 \pi^0 ) \over \Gamma(f_0(1500) 
\rightarrow \eta \eta)} = { p_{\pi} \over p_{\eta}} \left[ {1 \over
\cos^2(\phi)- \sin^2(\phi) } \right]^2   = 2.2
\ee
in fair agreement with the two experimental data
$1.45 \pm 0.61, \, 2.12 \pm 0.81$.

A little more complicate is the evaluation of $\Gamma(f_0(1500) \rightarrow 
\eta \eta ') / \Gamma(f_0(1500) 
\rightarrow \eta \eta) $ because $f_0(1500)$ just lies at the threshold
for $\eta \eta '$ production. One has to consider a weighting with
Breit-Wigner distribution through:

\bq
{\Gamma(f_0(1500) \rightarrow 
\eta \eta ') \over \Gamma(f_0(1500) 
\rightarrow \eta \eta)} = & \nonumber \\
2 \cdot \left [{2 \cos (\phi) \sin (\phi) 
\over \cos^2(\phi) - \sin^2(\phi)} \right ]^2 & \cdot \left[{
 \int_{m_{\eta}+m_{\eta '}}^\infty dE /[ ( 2 (E-M_{f_0}) /\Gamma ) ^2 +1] 
\cdot \sqrt{{(E^2-m_{\eta}^2 -m_{\eta '}^2)^2 - 4 m_{\eta}^2 m_{\eta '} ^2 
\over 4 E^4 }}    \over
{p_{\eta} \over M_{f_0}}  \int_0^\infty dE /[ ( 2 (E-M_{f_0}) 
/\Gamma ) ^2 +1]} \right ]\nonumber \\
= 0.232   \label{11}
\eq
to be compared with the value $0.29 \pm 0.10$, which is the Crystal Barrel result
reported in Ref. \cite{PDB}.

Analogously one has:
\bq
{\Gamma(f_0(1700) \rightarrow 
\eta \eta ') \over \Gamma(f_0(1700) 
\rightarrow K \bar K)} =&  \nonumber \\
1/2 \cdot \left [2 \cos (\phi) \sin (\phi) \cos^2(\alpha) \right ]^2 
\cdot & \left[{
 \int_{m_{\eta}+m_{\eta '}}^\infty dE /[ ( 2 (E-M_{f_0}) /\Gamma ) ^2 +1] \cdot \sqrt{{(E^2-m_{\eta}^2 
-m_{\eta '}^2)^2 - 4 m_{\eta}^2 m_{\eta '}^2 \over 4 E^4}}    \over
{p_{K} \over M_{f_0}}  \int_0^\infty dE /[ ( 2 (E-M_{f_0}) 
/\Gamma ) ^2 +1]} \right] \nonumber \\
= 0.039^{+0.013}_{-0.010} \, ,  
\label{12}
\eq
to be compared with future experimental data.

Further predictions may be obtained considering the decay of $J/\Psi$
($\Upsilon$) to
$f_0(1500)$ and $f_0(1700)$ to proceed mainly through the gluonic component of
these states. This assumption leads to:
\be
{\Gamma(J/\Psi \rightarrow \gamma f_0(1700) ) \over
\Gamma(J/\Psi \rightarrow \gamma f_0(1500)} = \tan ^2(\alpha) \cdot \left [
{M_{J/\Psi}^2 - 1.697^2 \over M_{J/\Psi}^2- 1.503^2} \right]^3= 0.39
^{+0.11}_{-0.14} \label{13},
\ee

\be
{\Gamma(\Upsilon \rightarrow \gamma f_0(1700) ) \over
\Gamma(\Upsilon \rightarrow \gamma f_0(1500)} = 0.50 
^{+0.14}_{-0.18} \label{14},
\ee

At the moment only the branching ratios $B.R.[J/\Psi \rightarrow \gamma
f_0(1700) \rightarrow \gamma K \bar K)] = (9.7 \pm 1.2) \cdot 10^{-4}$ and
$B.R.[J/\Psi \rightarrow \gamma f_0(1500) \rightarrow \gamma 4 \pi)] =
(8.2 \pm 1.5) \cdot 10^{-4}$ are available for the $J/\Psi$ decays and 
only an upper
limit for $f_0(1700)$ in the $\Upsilon$ case. 
Thus, due to the missing of the knowledge 
of $\Gamma(f_0(1500) \rightarrow 4 \pi)/ \Gamma(total)$, no real conclusion can
still be drawn.
When more stringent experimental
results will be available the ratios (\ref{13},\ref{14}) 
will represent a further test of this model.

Finally, another test of the model can be made considering the two--photon 
decays. Because gluons decouple from photons, this decay can
proceed only through the quark component and is therefore partially suppressed
for $f_0(1500)$ and $f_0(1700)$. In the mixing scheme (\ref{1}), one expects
\be
{\Gamma(f_0(1700) \rightarrow 
\gamma \gamma) \over \Gamma(f_0(1500) 
\rightarrow \gamma \gamma)}= \left({1.697 \over 1.503} \right) ^3 \cdot \cot ^2(\alpha) =
2.85^{+1.0}_{-0.8}
\label{2ph1}
\ee

If the $f_0(1400)$ is assumed to be the light quark isoscalar member of the
$0^{++}$ nonet (but this assignation is still quite controversial
\cite{PDB,Penn}) one also predicts
\be
{\Gamma(f_0(1700) \rightarrow 
\gamma \gamma) \over \Gamma(f_0(1400) 
\rightarrow \gamma \gamma)}= 0.095 ^{+0.016} _{-0.012}
\label{2ph2}
\ee
and
\be
{\Gamma(f_0(1500) \rightarrow 
\gamma \gamma) \over \Gamma(f_0(1400) 
\rightarrow \gamma \gamma)}= 0.033 ^{+0.008} _{-0.011}
\label{2ph3}
\ee

In summary, it has been reconsidered a very simple mixing scheme which enables
us 
to reproduce all the available experimental data (albeit still not very rich)
on the two glueball candidates $f_0(1500)$ and $f_0(1700)$. In this scheme both 
resonances have a gluonic component mixed with a $\bar s s$ one.

The  mixing angle has been obtained by a simple ansatz, allowing the
evaluation of several ratios of branching ratios of the two resonances, 
in good agreement with the available experimental data.

Using the predictions reported in this paper it will be possible to test 
with a larger accuracy the
model in a next
future, when new experimental results will appear, permitting a 
clearer understanding of the nature of these two peculiar particles and 
hopefully a first certain identification of a glueball.

Clarifying the nature of these two resonances will be of 
great help for understanding of the composition of the $0^{++}$ 
meson nonet, which is still quite controversial \cite{Penn,io2}.

\vskip 1.7cm
{\bf Acknowledgement}

Thanks are due to M. Anselmino and J.-M. Richard for useful comments.

\end{document}